\documentclass[global,twocolumn]{article}

\usepackage{epsfig} 
\usepackage{graphicx}

\usepackage[latin1]{inputenc}   % Use german keyboard / characters

\begin{document}

\title{Squeezed light from microstructured fibres: towards free space quantum cryptography}
\author{S. Lorenz\footnotemark[4] \and Ch. Silberhorn\footnotemark[4] \and N. Korolkova\footnotemark[4] \and R.S.
Windeler\footnotemark[5] \and G. Leuchs\footnotemark[4]\\
\footnotemark[4] Zentrum f\"ur moderne Optik, University Erlangen-N\"urnberg, Staudtstr. 7/B2, D-91058 Erlangen\\
\footnotemark[5] Bell Laboratories, Lucent Technologies, 700 Mountain Avenue, Murray Hill, New Jersey 07974\\
email: stefan.lorenz@physik.uni-erlangen.de 
}

\date{June, 30th 2001}

\maketitle
\abstract{Amplitude squeezed pulsed light has been produced using a microstructured silica fibre. By spectrally
filtering after the nonlinear propagation in the fibre a squeezing value of -1.7dB has been measured. A quantum
key distribution scheme based on 
squeezed light from such microstructured fibres is proposed.
}

\section*{Introduction}
Over the past years quantum effects have been demonstrated in systems described by continuous variables.
Entanglement was produced using nondegenerate parametric amplification \cite{OU92,FUR98,ZHA00}, Kerr effect in a silica
fibre \cite{SIL01} and four-wave mixing in
a fibre optic parametric oscillator \cite{SHA01}. One way to produce the entanglement is by interference of two squeezed
light beams. We introduce a new tool to generate non-classical light in the near infrared, which
shows great potential as compared to a device using standard silica fibres.

\section{Quantum effects in microstructured fibres}
The microstructured fibre evolved from experiments with photonic band gap materials.
It consists of pure silica, with a bulk core and a cladding which has a regular pattern of holes,
oriented in longitudinal direction \cite{RAN00}. Due to these holes, the cladding has a lower effective
refractive index than the solid silica core. This refractive index difference causes total internal reflection at 
the core-cladding
interface. The guiding effect thus is the same as in standard silica fibres.\\
However, the light mode is not completely confined in the core; there is a small
part of the mode which propagates in the cladding. Its relative portion depends on the wavelength. The effective
refractive index of the cladding therefore depends on the penetration depth, and thus
on the wavelenth of the propagating field.

\subsection{Pulse propagation}
Pulse propagation in fibers can be described by the so called non-linear Schrödinger-equation (NLSE)
\begin{equation}
i \frac{\partial A}{\partial z} - \frac{\beta_2}{2}\frac{\partial^2 A}{\partial T^2}+\gamma | A |^2 A=0,
\end{equation}
which describes the propagation of ps and sub-ps pulses \cite{AGR95}, neglecting higher order dispersion and higher order
nonlinearity. In case of a negative group velocity dispersion $\beta_2$ the fibre nonlinearity $\gamma$
allows for the 
formation of a stable pulse, called soliton \cite{HAS73,MOL80}. Due to the stability of the soliton against dispersive
broadening, the pulse's peak power remains virtually constant over a long interaction length. Thus,
even the low nonlinearity of silica can lead to significant quantum effects.

\subsection{Spectral filtering}
A stable soliton solution of the classical Equation (1) is a pulse with a special envelope
(hyperbolic-secant) and a particular peak amplitude. When quantizing the field the
corresponding solutions of the quantized version of Equation (1) are no longer stable, owing
to the amplitude and phase uncertainties. Thus the nonlinear Kerr effect induces new spectral sidebands in the pulse by self phase
modulation, depending on the amplitude of the propagating field, even for solitons. This
interaction between different spectral components of the pulse leads to amplitude noise
correlations. They can be exploited to reduce the overall noise in the pulse. This technique is known as squeezing by spectral filtering and
has been used for ps-pulses \cite{FRI96} and for sub-ps-pulses \cite{SPA97}. Numerical simulations
\cite{LEV99} as well as experimental tests \cite{SPA98b} have shown the high degree of correlation.
Our goal is to investigate whether the correlations also exist in the new microstructured fibre, where
solitons can propagate at a centre wavelength shorter than 800nm. In this interesting wavelength regime, only
non-soliton photon-number squeezing has been shown \cite{KON98,KRY99} so far.

\subsubsection{Experimental setup}
\begin{figure}
\includegraphics[width=\columnwidth]{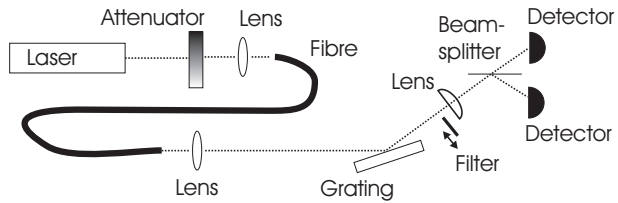}
\caption{\label{Bild_Setup}Experimental Setup}
\end{figure}

The experimental configuration is shown in Figure \ref{Bild_Setup}. 
The 130fs pulses of a commercial
Ti:Sapphire laser system (Spectra Tsunami, repetition rate 82MHz) were launched into a 1m piece of microstructured fibre
(Lucent) through a telescope and an aspheric lens. The output beam was collimated through another
aspheric lens and reflected by a blazed, gold plated grating (Jobin Yvon, 600/mm). The fibre end
was imaged to the filter plane, where a knife edge acted as a high pass filter. The filtered beam was then
focused by a cylindrical lens through a 50:50 beamsplitter onto two photodetectors which served as a
balanced detection systems (Photodiode: OSRAM BPW34, silicon PIN). The balanced detection enabled us to
measure amplitude fluctuations with reference to the shot noise level.
The sum and difference AC photocurrents of the two detectors were recorded by to electronic
spectrum analysers (HP 8590) at 23MHz with a bandwidth of 300kHz, a video bandwidth of 30Hz and averaged over 30
measurement cycles. 

\subsubsection{Results}
\begin{figure}
\includegraphics[width=\columnwidth]{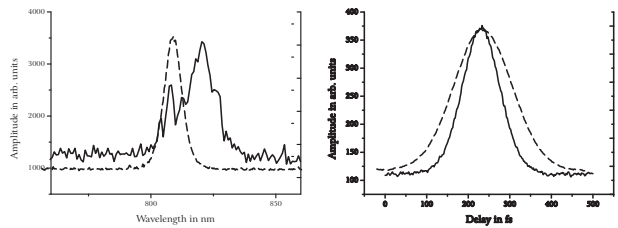}
\caption{\label{Bild_Spektren_AK}Left: Input and output spectra at an output power of 1.16mW corresponding to a
pulse energy of 14.1pJ. Right:
Corresponding autocorrelation traces. The dashed curves show the input pulse, the solid curves the output pulse
after the fibre.}
\end{figure}

\begin{figure}
\includegraphics[width=\columnwidth]{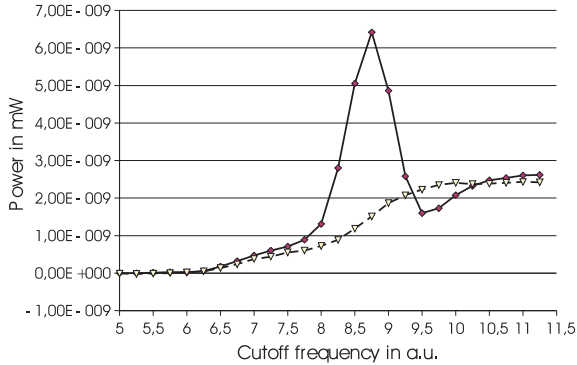}
\caption{\label{Bild_Squeezing}Typical measurement curve at a pulse energy of 15.8pJ. The AC noise powers of sum (solid line) and difference
(dashed line) 
photocurrents over the filter high pass cut-off are given in arbitrary units.}
\end{figure}

Figure \ref{Bild_Spektren_AK} shows typical in- and output spectra of the fibre, as well as
autocorrelation traces.
The nonlinearity of the microstructured fibre is about 20 times larger than that of standard telecommunication fibres
and
the soliton energy is 9pJ. For the measurement reported here a slightly higher pulse energy was used. This
explains the pulse narrowing shown in Figure 2. The pulse energy and the resulting average power is still ten times below saturation, in contrast
to previous experiments \cite{KON98}. In the
experiment reported here, the detector signal was only slightly above the electronic noise, so that the sum as well
as the difference photocurrent have been corrected for electronic noise of the detection system. Note that the
ratio between the optical signal or noise and the electronic noise will be an order of magnitude higher when using
heterodyne detection with local oscillator pulses as discussed below.\\
To check that the measured difference photocurrent corresponded to the shot noise, the beam was
attenuated with several neutral density filters. The variance of the photocurrent difference exhibited a linear
behaviour with respect to the beam power as expected for Poissonian statistics.\\
A typical measurement is shown in figure \ref{Bild_Squeezing}.
For a filter cut-off frequency between 9.5 and 10.25 (arbitrary units), the sum AC photocurrent drops below the difference photocurrent,
which represents the shot noise level. The maximum squeezing obtained is -1.7dB.

\subsection{Non-linear fibre Sagnac interferometer}
A squeezing source that uses spectral filtering has some inherent technical difficulties. The spectral
dispersion at the grating is very lossy, and the collimation of the filtered beam is tedious. There is 
also a fundamental limit for squeezing using spectral filtering \cite{MEC97}. We therefore
aim at another scheme of squeezed light production, which has proven to be useful before in a standard silica fibre \cite{SCHM98}. The
asymmetric nonlinear fibre Sagnac interferometer is known as a stable source for squeezed bright light pulses. The
input pulse is split into a bright and a dark pulse, which counter-propagate in the same fibre in the
interferometer. They interfere at the output, and for certain input energies they generate a bright amplitude
squeezed output pulse. Due to the insufficient polarisation maintaining properties of the fibre used, the double
NOLM principle (Nonlinear Optical Loop Mirror) \cite{SIL01} cannot be used. Thus, we plan to build two
independent NOLMs to produce two independently amplitude squeezed pulses.

\section{Possible entangled states}
With two amplitude squeezed beams, one can realise an entangled quantum state. In the past there have been
problems to define entanglement in the case of continuous variable systems. If one defines
entanglement as the non-separability of a state, necessary and sufficient conditions for entanglement of
discrete variable systems have been found by Peres \cite{PER96} and Horodecki \cite{HOR96,HOR97}. Their
criterion has been expanded by Duan et al. \cite{DUA00} and Simon \cite{SIM00} for the case of continuous variable systems.\\
If it comes to practical quantum communication, however, one might be interested in the correlations between
the two subsystems (A and B) of an entangled system. These are the correlations which help to infer the value of a variable measured at
subsystem A if the variable is already known at system B. This quantum mechanical property of an entangled pair was pointed out by
Einstein, Podolsky and Rosen \cite{EIN35} and later by Reid \cite{REI00} for optical fields. It is the basic building block of the proposed cryptography
system.

\subsection{Amplitude and phase entanglement}
The entanglement scheme applied with a quantum cryptographic application in mind uses uncertainty correlations
produced by the superposition of two
amplitude squeezed beams at a beamsplitter \cite{LEU99}. This results in amplitude "noise"
anti-correlations and phase "noise" correlations of the two output beams. The amplitude correlations have
been measured directly, the phase correlations indirectly for a NOLM made of a non polarisation preserving telecommunication
fibre \cite{SIL01}.

\subsubsection{Optical heterodyne detection}
The noise of a single beam can be measured by balanced optical heterodyne detection, using a strong 
local oscillator, a beamsplitter and a pair of balanced photodetectors. By variations of the phase of the
local oscillator, amplitude and phase noise of the signal beam can be measured. By now, an optical
heterodyne measurement has not been realised for intense beams, as the strong local oscillator would lead to saturation
effects in the photodetectors. A NOLM using microstructured fibres would operate at a lower signal power, so
that optical heterodyne detection may be used.

\subsection{Polarisation entanglement}
An alternative to amplitude and phase correlated beams is polarisation entanglement with intense beams
\cite{KOR01}. In contrast to dichotomic polarisation systems, not the polarisations themselves, but the 
variances of the Stoke's parameters are correlated. The detection of these correlations is possible by
use of passive, linear elements only, without the need for an local oscillator.

\section{Quantum cryptography}
Quantum cryptography uses quantum effects to establish a
random bit key, only known to two parties (Alice and Bob). The presence of an eavesdropper (Eve) can be
detected during the key generation.

\subsection{Free space quantum cryptography}
There are two applications which require free space transmission rather than fibre based communication.
The first is short distance communication up to several kilometers \cite{BUT98,RAR01}, mainly in urban areas, where a
fibre based connection is too expensive to deploy. The second is secure satellite communication, where a fibre
link is not possible.\\
For both applications a system operating at a wavelength of 800nm has several advantages over a system operating
at 1.5$\mu$m. The beam divergence
is considerably lower, allowing for smaller transmitter and receiver optics. The shorter wavelength also
allows for the use of silicon photodetectors, which are cheaper, have lower dark noise levels and are standard
industry products. In addition, the atmospheric transmission has several
transparency windows at 800nm, while there is considerable absorption at 1.5$\mu$m.

\subsection{A quantum key distribution scheme}
The quantum key distribution scheme we intend to use is described in \cite{SIL00c,SIL00b}. It depends on the correlations
that exist between amplitude and phase noise of an entangled pair of bright pulses. The bit value generated depends
on the type of measurement Bob and Alice perform: for example "1" for amplitude and "0" for phase measurement. Alice keeps one pulse of the
pair and sends the other to Bob over the quantum channel. They both now measure randomly and independently
amplitude or phase of their pulses, and Bob sends the results---but not the type---of his measurement back to Alice. If Alice can spot
correlations between her measurement and Bob's, she knows that they both have performed the same type of measurement. She
tells Bob that his measurement was correlated, and they each assign the appropriate bit values to theses measurement
slots.
If Alice detects a reduced degree of correlation, she knows that the quantum channel has been
manipulated.\\
We propose the implementation of the quantum key distribution system where the squeezed light
sources are built on the basis of NOLMs made of microstructured fibre. The scheme is secure against beamsplitting
attacks, provided the
correlations are strong enough. The speed of key generation is limited by the pulse repetition
rate of the system, as well as by the correlation measurement time. From the correlation measurements
already conducted we estimate the maximum achievable raw bit rate to be 10\% of the pulse repetiton rate. A
repetition rate of up to 100GHz is conceivable. 

\subsubsection{QKD for amplitude and phase correlations}
\begin{figure}
\includegraphics[width=\columnwidth]{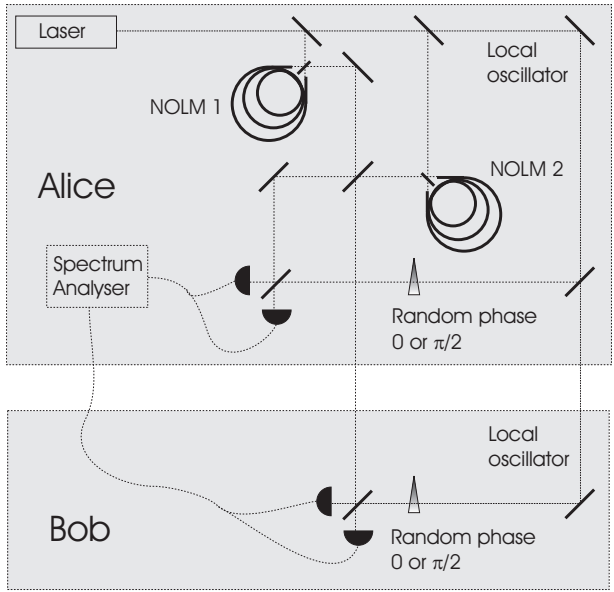}
\caption{\label{Bild_Setup_QKD}Proposed setup for the QKD system.}
\end{figure}

The proposed setup is shown in figure
\ref{Bild_Setup_QKD}.
Two independent NOLMs produce two amplitude squeezed light pulses, which are used to generate an
entangled pulse pair. The first pulse stays at Alice's site and either its phase or its amplitude
fluctuations are measured. The second pulse is transmitted to Bob, along with the local oscillator. Bob
also measures either amplitude or phase fluctuations and transmits his photocurrents back to Alice. She
checks the photocurrents for correlations and marks the measurements where she and Bob measured the same
variable. From the type of measurement Alice and Bob generate a string of shared random bits.

\subsubsection{QKD for polarisation correlations}
To implement a QKD-scheme using polarisation correlations as mentioned above, one would need four NOLMs, or two
with a polarisation maintaining fibre. The greater effort one has to put into the generation of the entangled pulses
is compensated by the substantially reduced complexity of the detection apparatus. There is no need to transmit a local
oscillator to Bob, so that his receiving station is a simple linear device with no need for interferometric
stability. We plan to update the amplitude/phase setup to a polarisation setup in a second step.

\section{Conclusions}
The potential of the microstructured fibre for the production of nonclassical bright light has
been shown. An amplitude squeezing of 1.7dB below shot noise was achieved using spectral
filtering. We hope to improve this value by optimizing system parameters such as the fibre length. A next goal is the implementation of the
microstructured fibre in two asymmetric fibre Sagnac interferometers, to produce an
EPR (Einstein Podolsky Rosen) entangled beam pair at 810nm. With the operation of the fibre based system in this new
wavelength regime free space quantum cryptography with continuous variables will become
viable.

\section{Acknowledgements}
The authors wish to thank Florence Gadaud, Oliver Glöckl, Michael Langer and Christoph Marquardt for their
support.\\
This work was supported by the german ministry of education and science BMBF under VDI-AZ 0155/00.
\bibliographystyle{unsrt}
\bibliography{berlin}

\end{document}